\documentclass[preprint,floats,superscriptaddress]{revtex4}

\usepackage{graphicx,color}
\usepackage[centertags]{amsmath}
\usepackage{tabularx}
\usepackage{delarray}
\usepackage{bm}
\usepackage[normalem]{ulem}
\usepackage{natbib}
\usepackage{url}

\begin{document}

\title{Shaped electric fields for fast optimal manipulation of electron spin
and position in a double quantum dot}

\date{today}

\author{J. A. Budagosky}
\affiliation{Institute for Biocomputation and Physics of Complex Systems,
  University of Zaragoza, Mariano Esquillor s/n, 50018 Zaragoza, Spain}

\author{D. V. Khomitsky}
\affiliation{Department of Physics, University of Nizhny Novgorod, 23 Gagarin
  Avenue, 603950 Nizhny Novgorod, Russian Federation}

\author{E. Ya. Sherman}
\affiliation{Department of Physical Chemistry, The University of the Basque
  Country UPV/EHU, 48080 Bilbao, Spain}
\affiliation{IKERBASQUE Basque Foundation for Science, Bilbao 48011 Bizkaia, Spain}

\author{Alberto Castro}
\affiliation{Institute for Biocomputation and Physics of Complex Systems,
  University of Zaragoza, Mariano Esquillor s/n, 50018 Zaragoza, Spain}
\affiliation{ARAID Foundation, Edificion CEEI, Mar{\'{\i}}a Luna 11,
  50018 Zaragoza, Spain}

\begin{abstract}
We use quantum optimal control theory algorithms to design external
electric fields that drive the coupled spin and orbital dynamics of an electron
in a double quantum dot, subject to the spin-orbit coupling and Zeeman
magnetic fields. We obtain time-profiles of multi-frequency electric field pulses which increase 
the rate of spin-flip transitions by several orders of magnitude in
comparison with monochromatic fields, where the spin Rabi oscillations were predicted to be very slow.
This precise (with fidelity higher than $10^{-4}$) and fast (at the timescale of the order of 0.1 ns, comparable with the Zeeman spin-rotation and the interdot tunneling time) simultaneous control of the spin and position is achieved while keeping the electron in the four lowest tunneling- and Zeeman-split levels through the duration of the pulse. Proposed optimized algorithms suggest effective  applications 
in spintronics and quantum information devices.
\end{abstract}


\maketitle

\section{Introduction}\label{introduction}

Among the current challenges in the fields of spintronics~\cite{Zutic2004} and
quantum information technologies based on semiconductor devices, one of the
most important is the fast and accurate manipulation of electron spin in
nanostructures and particularly in quantum dots (QDs)~\cite{Kouwenhoven2001}.
Since the electron spin is the most typical example of a two level
system, the possibility of using single electron QDs as a physical realization
of a \textit{qubit}~\cite{Loss1998} has been suggested and extensively studied.
Just very recently, the realization of a two-qubits logic
gate has been achieved in isotopically enriched silicon QDs\cite{Veldhorst2015}. 
Moreover, the spin-orbit coupling (SOC) -- present in semiconductors
used in manufacturing the QDs -- entangle the spin and the
orbital motion, making possible the handling~\cite{Rashba2003} and readout~\cite{LevitovRashba2003} 
of spin states fully by electric means. This ability of a monochromatic electric field
to coherently rotate the electron spin, demonstrated experimentally for
GaAs-based two-dimensional quantum dots in Ref.~[\onlinecite{Nowack2007}],
opened a venue for experimental and theoretical studies of spin manipulation
by electric fields. Several interesting regimes of the electric driving have been studied theoretically 
(see, e.g., Refs. [\onlinecite{Jiang2006,Bednarek2008,Rashba2011,Borhani2012,Nowak2012,Romhanyi2015,Osika2014}]) and 
algorithms for two-dimensional quantum dots to achieve the spin-flip faster than it can be done by
simple periodic fields, have been put forward \cite{Golovach2006,Budagosky2015,Ban2012}.

The recent progress in nanotechnologies, and the interest in novel systems
for semiconductors-based quantum technologies, have led to
the manufacturing of nanowire-based quantum dots \cite{Nadj2010}, and posed a
question about spin manipulation in these structures. The double quantum dots
can host two spatially separated electron spins, and thus constitute a prototype of
a basic two-qubits quantum information device. It is well-known that a single
isolated qubit cannot create a fully working quantum computational module, so
it is important to construct reasonable practical models of interacting qubits. 
A model of two spatially { separated} quantum dots coupled by tunneling can serve 
as a starting point. Here one deals both with the orbital degree of freedom describing
the electron location in a particular dot, and with the spin degree of freedom,
which can be coupled in a non-trivial way, for example, { by the}
SOC.  These considerations make the nanostructures with double QD interesting
candidates as hardware elements for future quantum information processing technologies.

However, even single-electron double quantum dots,  either two- or one-dimensional, 
where the electron can be localized
near one of the potential minima or delocalized in both of them, are qualitatively
different from single dots and can show a rather surprising spin 
dynamics (for two-dimensional dots see, e.g. Refs. [\onlinecite{Stano2006,Srinivasa2013}]).
First of all, these
two possible localizations can make a double quantum dot a realization of a
charge qubit.  Second, in contrast to a single dot, the low-energy
states in a double quantum dot are formed by interminima tunneling, making the
electron position strongly sensitive to any applied electric field that may
produce even a relatively weak asymmetry in the electron energies at the minima. 
Third, the electron motion
between the potential minima leads to a well-defined spin precession angle, proportional
to the tunneling distance. The second and the third factors make the
spin manipulation by electric fields in double and single quantum dots
qualitatively different. These differences can be summarized as follows: (i) The spin-flip Rabi frequency in a double quantum dot becomes much lower than expected from the conventional linear dependence on the driving field amplitude, and (ii) the involvement of several
low-energy orbital states produces a mixed rather than a pure
spin state, pushing the spin vector inside the Bloch sphere
\cite{Khomitsky2012}.  (For a one-dimensional single quantum dot see Ref. [\onlinecite{Li2013}].) 
As a result, a full spin-flip is difficult to achieve
even at long times, thus, making the spin manipulation strongly sensitive to the decoherence.

To eliminate these obstacles and speed-up the spin manipulation, 
one may try to apply a modification of the driving technique, namely, to
control the quantum dynamics via
specially tailored electric fields, rather than using monochromatic
ones. The theoretical design of
the shape of the fields can be achieved with the help of quantum optimal
control theory \cite{Werschnik2007, Brif2010} (QOCT). Here, we employ this
technique to study the possibility of spin-flip speed-up, computing the 
fields that allow to produce the requested spin states and electron
displacements in a fast and controllable way.

This paper is organized as follows. In Sec.~\ref{model}, we introduce the
quantum mechanical model of a double quantum dot (DQD), hosting one 
electron, in the presence of SOC and Zeeman magnetic fields, and a
time-dependent external electric field that can drive the dynamics.  In
Sec.~\ref{qoct}, we describe our implementation of QOCT for this system. In
Sec.~\ref{results}, we present relevant results, that show, in particular, how
the spin-flip rate can be strongly increased in comparison to the rates computed 
for a monochromatic driving field in a
previous study on the same system~\cite{Khomitsky2012}. The
conclusions of this work are given in Sec.~\ref{conclusions}.

\section{Model}\label{model}

\begin{figure}
\includegraphics[trim=0cm 0cm 0cm 0cm, clip=true,width=8cm]{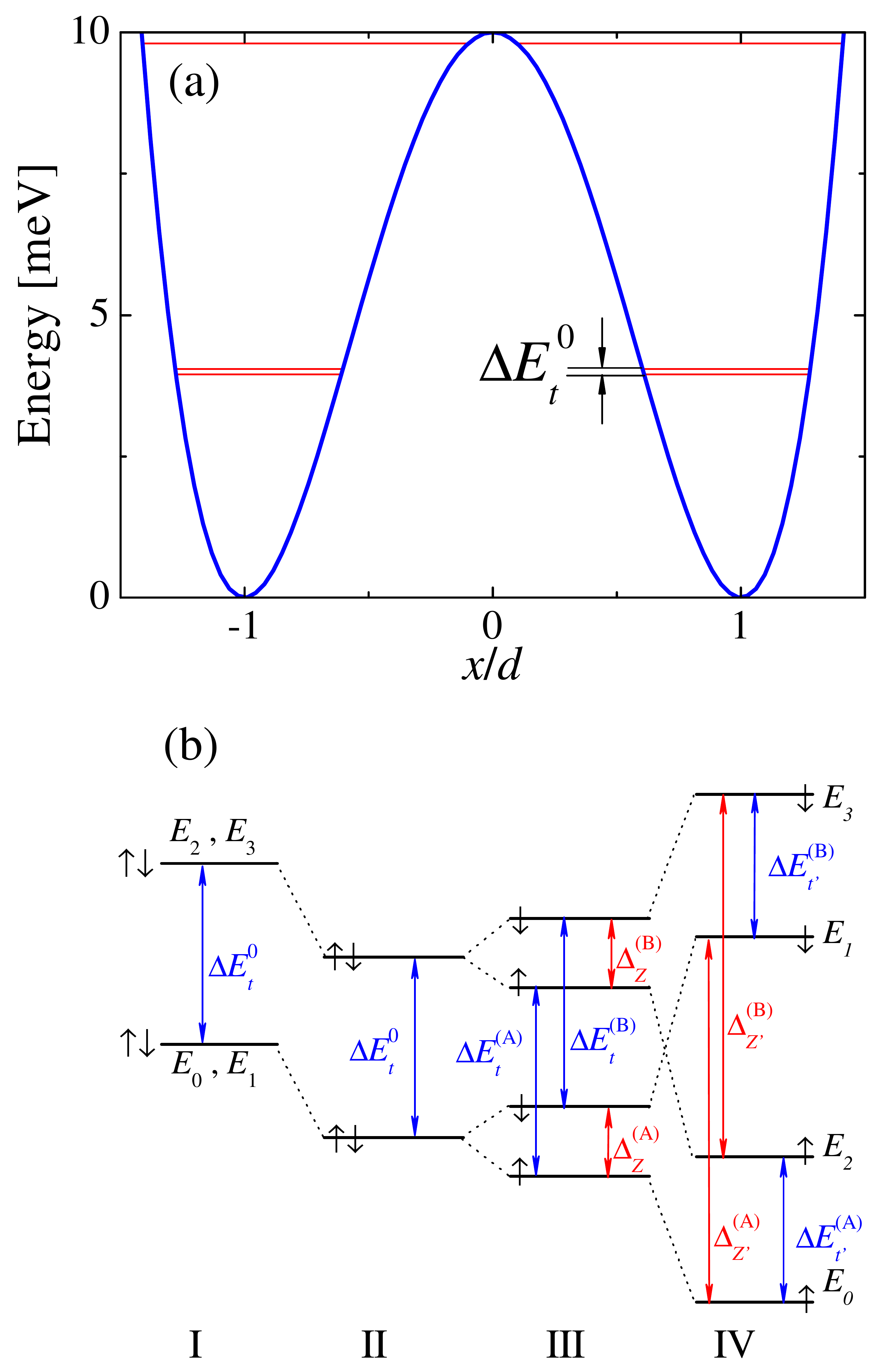}\\
\caption{ (Color on-line) (a) Confinement potential $U(x)$ and the six first
  energy levels of the DQD system with the SOC and magnetic field
  switched-off. Each horizontal red line is two-fold degenerate in spin. (b)
  The four lowest energy levels for four cases: (I) without SOC and magnetic
  field; with the SOC switched-on ($\alpha_R=10^{-9}$ eV cm and
  $\alpha_D=0.3\times 10^{-9}$ eV cm) for (II) $B=0$ T, (III) $B=1.73$ T and
  (IV) $B=6.93$ T.  }
\label{fig-1}
\end{figure}

We use the model of Ref. [\onlinecite{Khomitsky2012}], which can be realized by producing a nanowire-based 
double quantum dot with the gating technique described, for example, in Ref. [\onlinecite{Nadj2012}].
In the absence of SOC and external magnetic field, the Hamiltonian
\begin{equation}\label{eq-1}
\hat{H}_k=\frac{k^{2}}{2m^{*}}+U(x)
\end{equation}
describes an electron confined in a one-dimensional double quantum
dot, where $k=-i\partial/\partial{x}$ is the momentum operator (we set $\hbar\equiv 1$),
$m^{*}$ is the electron effective mass, and 
the confinement potential $U(x)$ is given by:
\begin{equation}\label{eq-2}
U(x)=U_0\left[ 1 + \left(\frac{x}{d}\right)^4 - 2\left(\frac{x}{d}\right)^2 \right]\,.
\end{equation} 
The shape of this potential is presented in Fig.~\ref{fig-1}. It is a
double-well structure, with two minima at $-d$ and $d$ ($U(d)=U(-d)=0$),
separated by a barrier of height $U_0$. A characteristic tunneling time $T^{0}_{t}=2\pi/\Delta E^0_t$ is
determined by the tunneling energy $\Delta E^0_t \ll U_0 $, that is the gap
between the ground and first excited state of $\hat{H}_k$. 

The presence of an external magnetic field may then be accounted for by a Zeeman coupling term
(the vector potential term does not enter the Hamiltonian Eq.~\eqref{eq-1} for a 1D structure),
defined by $\hat{H}_Z=\Delta^0_Z\hat{\sigma}_z/2$, where $\Delta^{0}_{Z}=g^*\mu_{B}B$,
and $\hat{\sigma}_z$ is the corresponding Pauli matrix (being $g^*$ and $\mu_B$ the effective $g$-factor and the
Bohr magneton, respectively). In addition there is a SOC term in the form:
\begin{equation}\label{eq-3}
\hat{H}_{\text{SO}}=\left( \alpha_D\hat{\sigma}_x + \alpha_R\hat{\sigma}_y \right)k~\text{.}
\end{equation}
The strength of the SOC is determined by the structure-related Rashba
($\alpha_R$) and the bulk-originated Dresselhaus ($\alpha_D$) parameters.

Finally, under the influence of a driving electric field, the static Hamiltonian 
\begin{equation}
\hat{H}_0 = \hat{H}_k + \hat{H}_{\rm SO} + \hat{H}_Z
\end{equation}
described above is supplemented with a time-dependent term, and the system is governed by
the time-dependent Schr\"{o}dinger's equation during the time interval $[0,
  T]$:
\begin{equation}\label{eq-4}
i\frac{d}{dt}\vert \Psi(t)\rangle=\hat{H}(t)\vert\Psi(t)\rangle = 
\left[ \hat{H}_0 + e\hat{x}\varepsilon(t) \right]\vert \Psi(t)\rangle\,,
\end{equation}
where the electron-field interaction assumes the
dipole-approximation in the length gauge. 

Numerically, we first employ a truncated basis of planewaves, i.e. assuming a periodic unit cell of size $L\approx 7d$ (large enough to ensure convergence) as simulation box:
\begin{equation}
\langle x \vert j \rangle = \frac{1}{\sqrt{2\pi L}
}e^{i\frac{2\pi j}{L}x}\,.
\end{equation}
This basis is used to represent
the time-independent Hamiltonian $\hat{H}_0$:
\begin{equation}
H^0_{jj'} = \langle j \vert \hat{H}_0 \vert j'\rangle\,.
\end{equation}
Then, we diagonalize the resulting matrix $H^0_{jj'}$, 
obtaining the eigenbasis $\lbrace \psi_n\rbrace$:
$\hat{H}_0\vert\psi_n\rangle
= E_n \vert\psi_n\rangle$. Later, in order to solve Eq.~\eqref{eq-4},
the wave function $\Psi(t)$ is expanded in this basis:
\begin{equation}\label{eq-5}
\Psi(x,t)=\sum_m\zeta_m(t)e^{-iE_m t}\langle x|\psi_m\rangle~\text{.}
\end{equation}
Substituting Eq.~\eqref{eq-5} in~\eqref{eq-4}, we obtain for the time dependence of $\zeta_m(t)$: 
\begin{equation}\label{eq-6}
\frac{d}{dt}\zeta_n(t)=ie\varepsilon(t)\sum_m\zeta_m(t)\langle \psi_n |\hat{x} | \psi_m \rangle e^{i(E_n-E_m)t}~\text{.}
\end{equation}

\section{Quantum optimal control theory}\label{qoct}

The evolution of the system modeled in the previous section is determined
by the external field $\varepsilon(t)$. The goal now is to shape
this real function in order to bring the system to in a pre-defined state, for
example, switching the spin direction. As we mentioned in
Sec.~\ref{introduction}, the theoretical tool suitable for this task is
QOCT. It has already been used recently to study the electron dynamics
of 2D quantum dots and rings in the presence of THz laser
fields~\cite{Rasanen2007,Rasanen2008a,Rasanen2008b,Budagosky2015}.

The formalism of QOCT for the purposes of this work can be summarized as
follows: the function $\varepsilon(t)$ (hereafter, the \emph{control function})
is determined by a set of parameters $u_1,\ldots, u_M\equiv\bm{u}$:
$\varepsilon(t)\equiv\varepsilon[\bm{u}](t)$. The specification of a
particular set $\bm{u}$ therefore fully determines the evolution of the system:
$\Psi = \Psi[\bm{u}]$. Next, one must encode the physical target that one
wants to achieve into the mathematical definition of a \emph{target functional} $F$,
that depends on the system state $\psi$, and possibly also on the control
function parameters: $F=F[\Psi,\bm{u}]$. The value of this functional
determines the degree of success achieved: a large value of $F$ should
correspond to a  desirable realization of the physical target.

In our case this functional is split into two parts,
$F[\Psi,\bm{u}]=J_1[\Psi]+J_2[\bm{u}]$, where $J_1$ depends on the state of
the system only and $J_2$ is determined solely by the control function. Regarding $J_1$, 
it may depend on the evolution of
the system during all the propagation time, or only on its final state:
$J_1[\Psi]=J_1[\Psi(T)]$, as it is assumed in this work. It is typically
defined as the expectation value of some operator $\hat{O}$:
\begin{equation}\label{eq-8}
J_1[\Psi]=\langle\Psi(T)|\hat{O}|\Psi(T)\rangle\,.
\end{equation}

Regarding $J_2$, sometimes called the \emph{penalty}, it may be 
included in the definition of $F$ in order to
\emph{penalize} { entering} unwanted regions of the search space; for example, we wish 
to avoid solutions with high intensities, and this can be done by defining
$J_2$ as:
\begin{equation}\label{eq-9}
J_2[\bm{u}]=- \gamma \int_0^T\!\!\varepsilon^2[\bm{u}](t)dt~\text{,}
\end{equation}
where $\gamma$ is a positive constant, the \emph{penalty factor}. In this way,
as the fluency (integrated intensity of the driving field) is higher, the value of $J_2$
is negative and larger in the absolute value and the sum $J_1+J_2$ is smaller.
Thus, the  $J_2-$term ensures that the optimization algorithm will try to find
solutions simultaneously maximizing the physical target and minimizing the pulse fluency.

Since the shape of the control function and the evolution of the system
is determined by the parameters $\bm{u}$, the entire problem is
reduced to the maximization of a function of $\bm{u}$:
\begin{equation}\label{eq-10}
G[\bm{u}]=F[\Psi[\bm{u}],\bm{u}]=J_1[\Psi[\bm{u}]]+J_2[\bm{u}].
\end{equation}

There are many alternative algorithms to maximize a real function of many
arguments such as $G$; the most effective ones necessitate procedures
to compute the value of both the function and its gradient.
With the QOCT approach one obtains for the gradient of $G$:
\begin{eqnarray}
\nonumber
\nabla_{\bm{u}}G[\bm{u}] & = & 2{\rm Im}\left[\int_0^T dt 
\langle\chi[\bm{u}](t)|\nabla_{\bm{u}}
\hat{H}[\bm{u}](t)|\Psi[\bm{u}](t)\rangle\right]
\\\label{eq-11}
& & + \nabla_{\bf u} J_2[{\bf u}]\,,
\end{eqnarray}  
where, given the structure of our Hamiltonian, $\nabla_{\bm{u}}
\hat{H}[\bm{u}](t)=\hat{x}\nabla_{\bm{u}}
\varepsilon[\bm{u}](t)$. Here, $\chi[\bm{u}](t)$ is a new auxiliary backward propagation wave function, defined as the solution of:
\begin{subequations}
\begin{eqnarray}
i\frac{d}{dt}\chi[\bm{u}](x,t)&=&\hat{H}^{\dag}[\bm{u}](t)\chi[\bm{u}](x,t)\,,\label{eq-12a}\\
\chi[\bm{u}](x,T)&=&\hat{O}\Psi[\bm{u}](x,T)\label{eq-12b}\,.
\end{eqnarray}
\end{subequations}
The maxima of $G$ are found at the critical
points $\nabla_{\bm{u}}G=0$ by using the quasi-Newton method 
designed by Broyden \textit{et al}.~\cite{BROYDEN01031970,Fletcher2000}.

Regarding the target operator $\hat{O}$, its definition depends on the goal,
which in this work is always to populate some selected excited state 
$\psi_t$, that has the required spin (orbital)
orientation (position). Obviously, one easy way to achieve this goal is to use
the projection onto that state and choose $\hat{O}=|\psi_t\rangle\langle\psi_t|$. Thus,
the functional $J_1$ has the form:
\begin{equation}\label{eq-13}
J_1[\Psi]=\langle\Psi(T)|\hat{O}|\Psi(T)\rangle=|\langle\psi_t|\Psi(T)\rangle|^2.
\end{equation}
The initial and target states can be chosen as 
linear combination of eigenstates of the system. 

It remains to specify the parametrization of the control function,
i.e. the definition of the parameter set $\bm{u}$.  In our case, we expand the
control function in a Fourier series, so that the  parameters correspond to
the coefficients. However, this correspondence is not exact, since 
we must enforce several physical constraints. First, in order to ensure that the signal
over the full propagation time
integrates to zero, the zero-frequency component is assumed to be zero. Second, in
order to ensure that the field starts and ends at zero, the sum of all the
cosine coefficients of the Fourier series is also set to zero.

Finally, since these optimizations are iterative algorithms, one must assume an
initial driving field. Thus, in all the cases discussed below, we start by
considering a ``reference'' field of the form:
\begin{equation}\label{eq-14}
\varepsilon_{\text{ref}}(t)=A_0\sin(\omega_0t),
\end{equation}
where $\omega_0$ is the characteristic frequency of the ``target transition'',
i.e. the difference in energies between the initial and target states (the
expectation values whenever those states are not eigenstates), and $A_{0}$ is the amplitude.

\section{Results}\label{results}

In the following, we present results for three 
types of transitions in a double quantum dot. 
Specifically, the first two examples show the realization of a
spin-flip transition for two different external magnetic fields. In the
third example we show the possibility of controlling the electron position,
moving it from one minimum of the DQD { potential} to the
other. In the last example we explore the simultaneous control of spin
and position.

For all our calculations we have considered a GaAs-based structure,
with an effective mass of $m^*=0.067m_e$ ($m_e$ is the free electron mass) and
$g^*=-0.45$. The potential minima are set at $d=25\sqrt{2}$ nm and $U_0=10$ meV (unless otherwise indicated). These
parameters are the same as those considered in
Ref.~[\onlinecite{Khomitsky2012}]. This will allow us to compare some of the new
results with those of that work. In the absence of magnetic field and SOC, the
tunneling splitting is $\Delta E^0_t=0.0928$ meV (see Fig.~\ref{fig-1}).  This
value corresponds to a transition frequency of approximately $23$ GHz. Note
also in Fig.~\ref{fig-1}(a) that the lowest tunneling-split doublet is separated 
by a large ($\approx 6$ meV) gap from the higher orbital states. As discussed below, 
this aspect is of great importance in relation to the fidelity of the achievable spin states. 
The presence of SOC does not modify $\Delta E^0_t$, and the most
noticeable effect is a global redshift, as it can be learned from case II in
Fig.~\ref{fig-1}(b).

The presence of the SOC alone does not break the time reversal symmetry of the 
electron states. Therefore, the spin twofold degeneration is preserved unless
we introduce an external magnetic field, as can be seen in cases III and IV in
Fig.~\ref{fig-1}(b), that correspond to a moderate ($1.73$ T) and strong
($6.93$ T) magnetic fields. If the SOC is not present, these 
magnetic fields correspond to Zeeman splittings $\Delta^0_Z=\Delta E^0_t/2$ and
$\Delta^0_{Z'}=2\Delta E^0_t$, respectively. We may then define other
characteristic times and energies of the system: the Zeeman splittings
$\Delta^{\rm (A)}_{Z}=0.0358$ meV and $\Delta^{\rm (B)}_{Z}=0.0355$ meV (and
the corresponding periods $T^{\rm (A)}_{Z}=2\pi/\Delta^{\rm
  (A)}_{Z}=115.536$ ps and $T^{\rm (B)}_{Z}=2\pi/\Delta^{\rm
  (B)}_{Z}=116.589$ ps) at the low magnetic field (1.73 T) and the
Zeeman splittings $\Delta^{\rm (A)}_{Z'}=0.18735$ meV and $\Delta^{\rm
  (B)}_{Z'}=0.18606$ meV, associated to the higher magnetic field (6.93 T)
($T^{\rm (A)}_{Z'}=2\pi/\Delta^{\rm (A)}_{Z'}=22.074$ ps and $T^{\rm
  (B)}_{Z'}=2\pi/\Delta^{\rm (B)}_{Z'}=22.228$ ps). It is important
to mention that in the absence of SOC $\Delta^{\rm (A)}_{Z}=\Delta^{\rm
  (B)}_{Z}=\Delta_{Z}$ and $\Delta^{\rm (A)}_{Z'}=\Delta^{\rm
  (B)}_{Z'}=\Delta_{Z'}$.

Unless otherwise indicated, the peak amplitude of the reference field is set to
$1.5\times10^2$ V$/$cm, which corresponds to
$2d\times eA_0\approx 0.1U_0\approx 10\Delta E_{t}^{0}$, strongly modifying the 
interminima tunneling.
The latter estimate means that the field is strong enough 
to potentially involve also the states higher than the four states
shown in Fig.~\ref{fig-1}.
To calculate the dynamics, we have represented the
wavefunction using a basis of 20 eigenstates (see Eqs.~\eqref{eq-5}
and ~\eqref{eq-6}). Finally, for the parameterization of the electric field in
Fourier series, we have set in all the examples below a cut-off frequency
$\omega_{\rm{cut-off}}=10\omega_0$.

\subsection{Control of spin-flip dynamics}\label{cdsfd}


For the first example we consider a moderate external magnetic field,
$B=1.73$ T with the chosen Rashba and Dresselhaus SOC parameters being the same as
introduced earlier. We start with a reference field whose driving frequency
$\omega_0=\Delta^{\rm (A)}_Z$ is in resonance with the transition that we want to maximize, that is between the states $\psi_0$ and $\psi_1$:

\begin{equation}\label{eq-15}
\psi_0 \rightarrow \psi_1
\end{equation}

\begin{figure}
\centering
\includegraphics[trim=0cm 0cm 0cm 0cm,clip=true,width=0.425\textwidth]{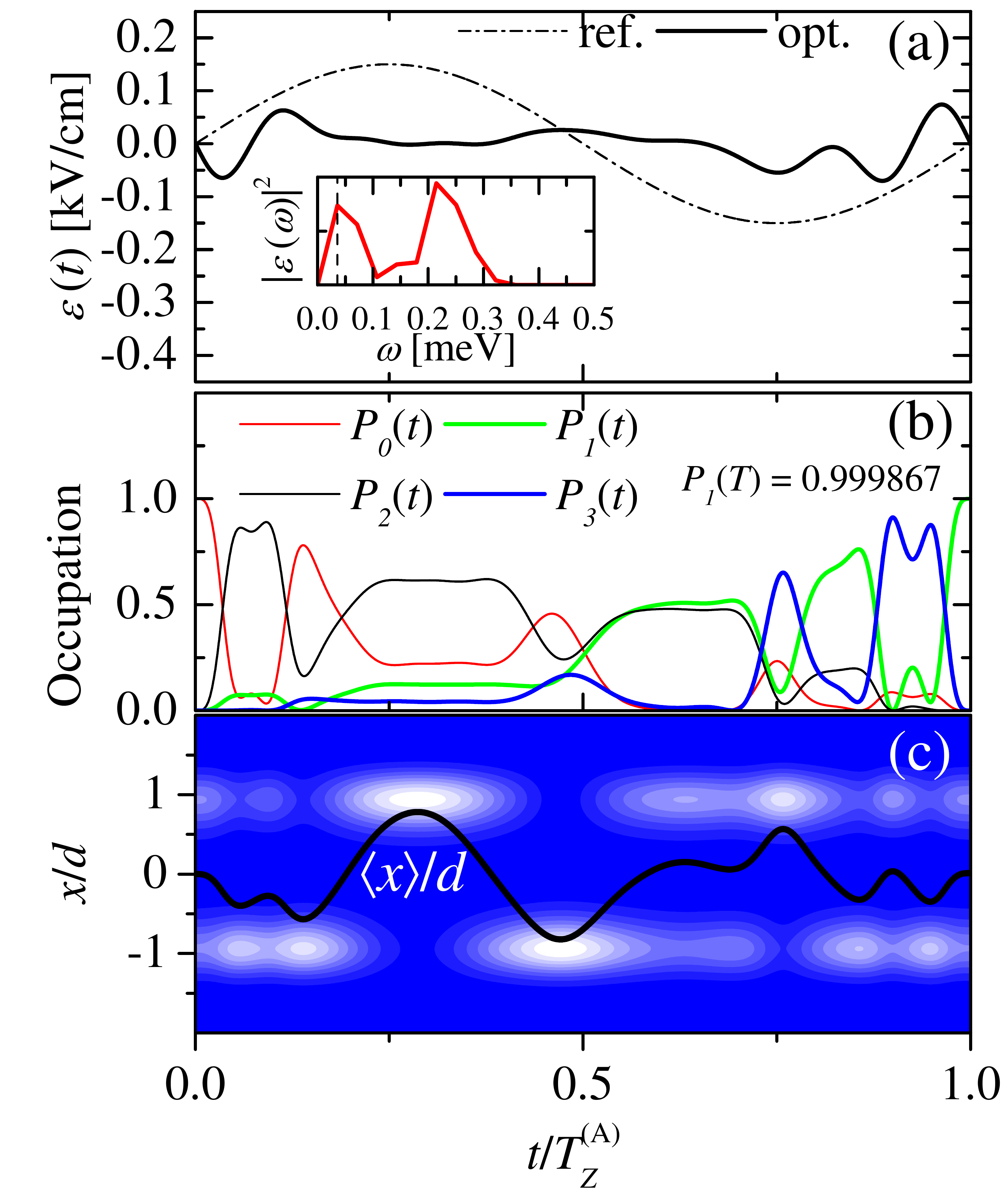}\\
\includegraphics[trim=0cm 0cm 0cm 0cm,clip=true,width=0.425\textwidth]{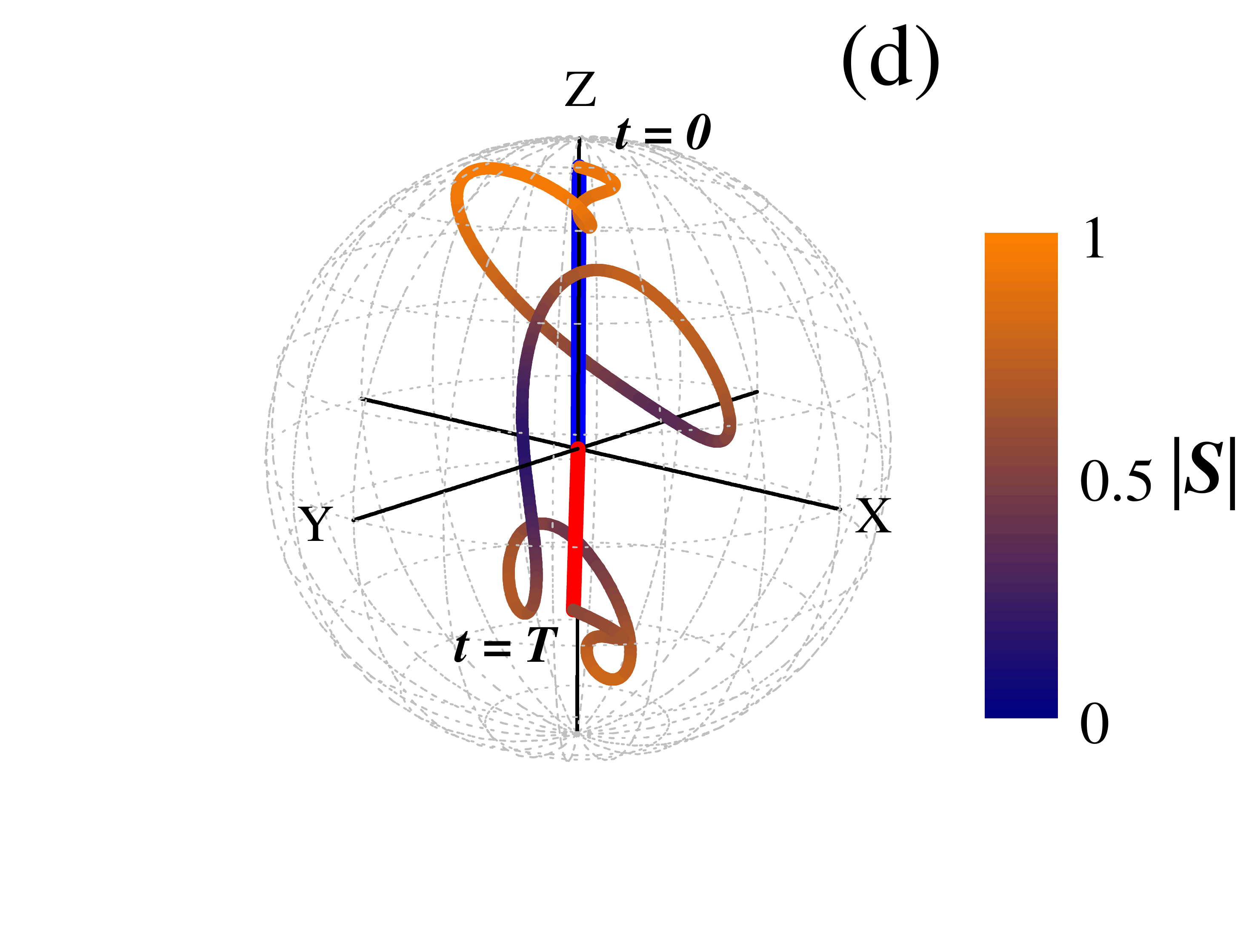}\\
\caption{(Color on-line) (a) Optimized driving field for a maximized spin-flip
  transition between states $\psi_0$ and $\psi_1$ with a magnetic field of
  $B=1.73$ T (the reference driving field is also shown for comparison). In
  the inset, the power spectrum (arbitrary units) in frequency domain of the optimized pulse is
  shown. The vertical dashed line marks the initial driving frequency; (b)
  Time-dependent occupation of the four lowest energy levels; (c) time-dependence of the charge distribution. 
  The white color refers to the maximum values 
  of charge. The thick black line is the time-dependent $\langle x\rangle$; 
  (d) trajectory of the tip of the spin vector, ${\bm S}(t)$ in the Bloch sphere. 
  The color scale represents the module of ${\bm S}$. The blue and red thick lines 
  indicate the initial and final spin vectors, respectively.
} \label{fig-2}
\end{figure}

The Figure~\ref{fig-2}(a) shows the electric field pulse resulting from the
optimization, and the thin line is the reference initial field
$\varepsilon_{\text{ref}}(t)$, shown for comparison.  The time scale is in the 
units of $T^{\rm (A)}_Z$.  Note that the optimized pulse has a lower fluency
than the reference field, and contains higher-frequency
oscillations, forming a very irregular temporal profile. These
higher-frequencies components are better observed in the power spectrum in the
inset in Fig.~\ref{fig-2}(a), where one can see a bimodal frequency
distribution. A sharp distribution of frequencies is centered around the
initial $\omega_0=\Delta^{\rm (A)}_Z$, forming the lowest energy
peak.

This optimized pulse produces the dynamics of the population of the
levels shown in Fig.~\ref{fig-2}(b). In this plot, only the first
four levels are shown since, throughout all the pulse duration, the electron
occupies mainly { these four states}.  Clearly, in this case the target
is achieved almost entirely ($0.999867$, as indicated in the plot). As it was
previously discussed in Ref.~[\onlinecite{Khomitsky2012}], the dynamics of this
system can be characterized approximately as a superposition of two kinds of
transitions: those of spin-flip type and those of spin-conserved type. This
aspect can be clearly seen in Fig.~\ref{fig-2}(b), where one can notice the
presence of fast exchanges between the populations of the states $\psi_0$ and
$\psi_2$, at the beginning of the pulse, and between the populations of the
states $\psi_1$ and $\psi_3$ at the end of the pulse.  These two 
{ spin-conserving transitions} are related with the tunneling splittings $\Delta
E^{\rm (A)}_t$ and $\Delta E^{\rm (B)}_t$, respectively. 
As we mentioned
earlier, the transition that we want to maximize (Eq.~\ref{eq-15})
is of spin-flip type, and is associated to the spin splitting $\Delta^{\rm (A)}_Z$, 
which is smaller than the tunneling. In contrast with the fast
spin-conserved transitions, note the slower general depopulation (population)
of states $\psi_0$ and $\psi_2$ ($\psi_1$ and $\psi_3$). Note also in Fig.~\ref{fig-2}(c) how the average position of the electron smoothly moves 
from one QD to the other (the charge distribution varies slowly between minima $d$ and $-d$) in time intervals where the occupation of the states changes
very little--or does so less abruptly. In contrast, note that at the beginning and at the end of the interval $[0,T]$ 
the charge is distributed almost equally 
between both QDs, showing just faster and smaller oscillations. This explains why in these 
regions the average position of the electron is found mainly 
within the region of the tunneling barrier. Note also that when only one eigenstate is 
occupied, the charge is equally distributed in both QDs, while a charge 
located mainly in one of these involves a superposition of two or more states.

Finally, Fig.~\ref{fig-2}(d) shows the time evolution of the spin vector inside the 
Bloch's sphere. Since the target eigenstate $\psi_1$ is almost completely
populated, it is therefore not surprising that the final value of
$\langle\sigma_z\rangle$ corresponds to that eigenstate. Due to the
SOC, electron spin is no longer well defined, so the $z-$component
does not reach $\langle\sigma_z\rangle=-1$. Note also
the absence of high-frequency oscillations in the spin vector evolution.
These rapid oscillations are associated with transitions involving higher
energy levels, so their absence reveals that the occupation of the higher orbitals
during the evolution time is negligible.

\begin{figure}
\centering
\includegraphics[trim=0cm 0cm 0cm 0cm,clip=true,width=0.425\textwidth]{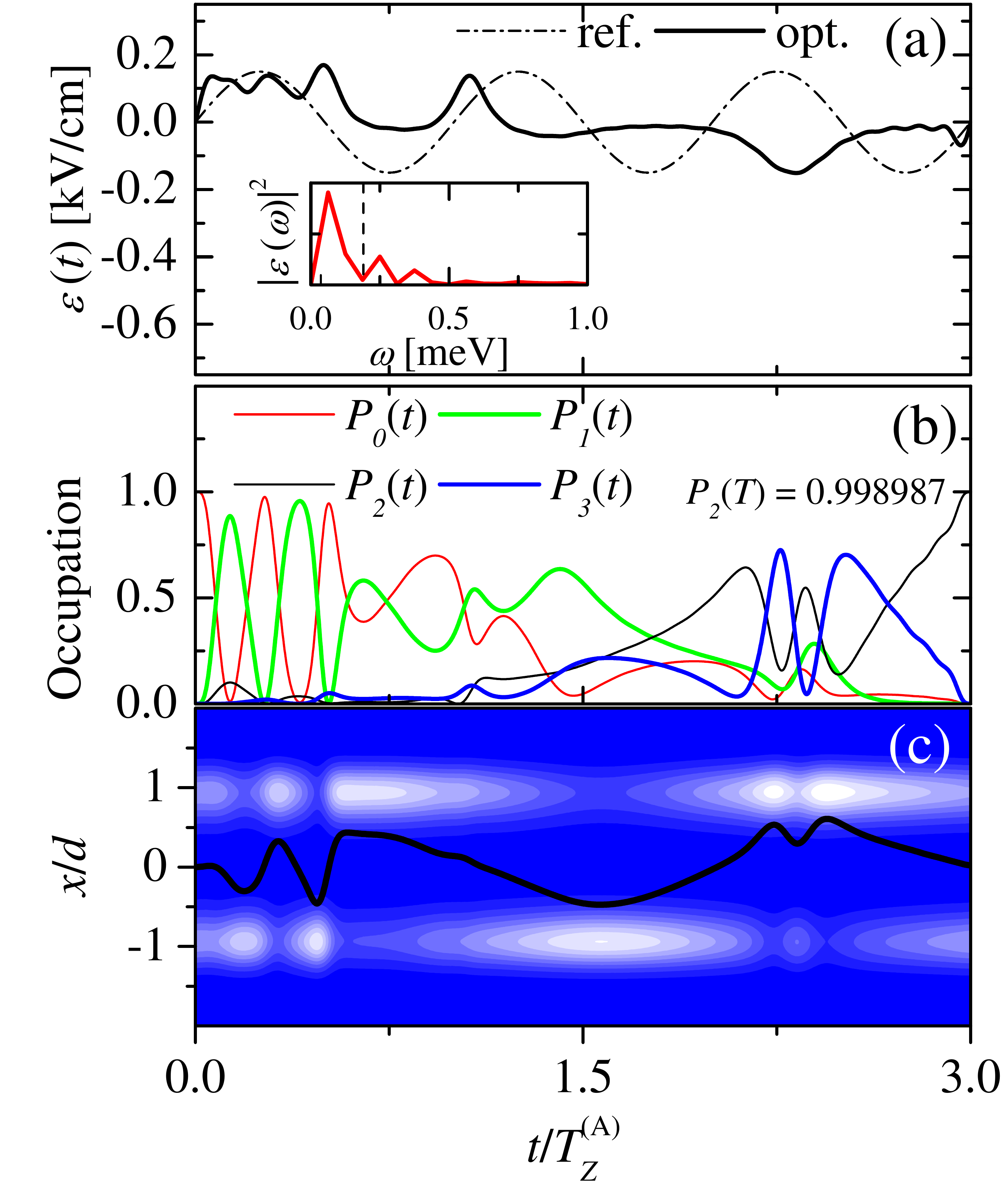}\\
\includegraphics[trim=0cm 0cm 0cm 0cm,clip=true,width=0.425\textwidth]{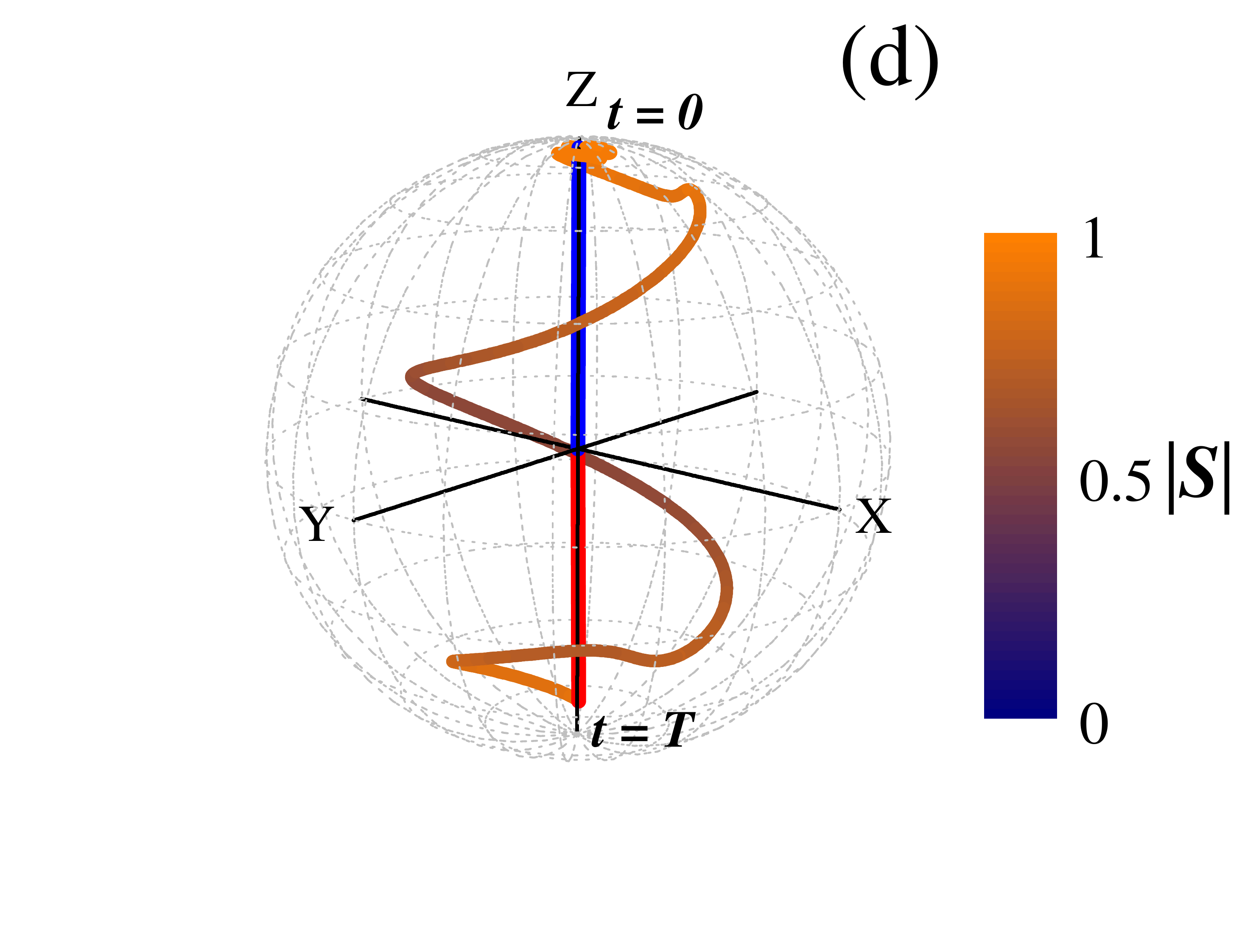}\\
\caption{(Color on-line) The same as Fig.~\ref{fig-2} but with a magnetic field of $B=6.93$ T. 
In this case, the target is the state $\psi_2$.} \label{fig-3}
\end{figure}


Next, we have considered the same spin-flip goal, but using a relatively strong
external magnetic field, $B=6.93$ T.  Here, energy levels $E_0$ and $E_1$
correspond both to ``spin-up'' states, while levels $E_2$ and $E_3$ correspond
to ``spin-down'' states.  In this case, the goal is to maximize the 
transition between states $\psi_0$ and $\psi_2$ ($\omega_0=\Delta^{\rm
(A)}_{Z'}$):

\begin{equation}\label{eq-16}
\psi_0 \rightarrow \psi_2
\end{equation}

The results are shown in Figs~\ref{fig-3}.  Importantly here, the
precession time, $T^{\rm (A)}_{Z'}$, associated with the target transition is
smaller than in the previous case since this is a higher energy transition.  As a
consequence it has been possible to achieve { good} results in terms of
the target only by setting $T=3T^{\rm (A)}_{Z'}=66.223$ ps as the minimum
duration for the pulse.

In general terms, the behavior observed in this regime of stronger magnetic
field is similar to the behavior typical for a moderate $B$. 
However, in this case the optimized { pulse exhibits peaks that
are slightly higher than the reference field,} as it can be seen
in Fig.~\ref{fig-3}(a). This has consequences for the occupation of higher
energy levels, as discussed below. In the inset of Fig.~\ref{fig-3}(a), we
note that the distribution of the frequencies that comprise the optimized
pulse is broader than that for a moderate magnetic field. In addition,
there is only one main peak, centered around a frequency equal to
$\Delta^{\rm (A)}_{Z'}/3$. Furthermore, in Fig.~\ref{fig-3}(b) one can observe at the
beginning (end) of the time interval, a strong exchange of population between
levels $E_0$ and $E_2$ ($E_1$ and $E_3$), that is, between pairs of states
with the same spin orientation. { As in the previous case, the dynamics of the average position 
of the electron in Fig.~\ref{fig-3}(c) shows a higher 
frequency oscillatory behavior of $\langle x \rangle$ in the intervals of fast spin-conserved 
transitions, whereas in the rest of the time interval 
the dynamics is less abrupt, showing a slow oscillatory displacement between both minima, $d$ and $-d$}. 
The result of the optimization can be 
considered as satisfactory given the degree of occupation of the target at the end of the pulse as 
well as the final spin orientation (Fig.~\ref{fig-3}(d)).

In these two examples we have demonstrated how { electric field pulses}
with suitably designed temporal profiles can produce spin-flip transitions on
time scales much shorter than the period { of} the corresponding Rabi
oscillations: these periods were shown in Ref.~[\onlinecite{Khomitsky2012}] to
be about one order of magnitude longer.

It becomes therefore clear that the tailored shapes allow for population
transfer mechanisms that cannot be accessed with the simple Rabi-like
resonance-based processes. For example, in these two first cases the
population dynamics has three phases. The first phase is characterized by a
fast exchange between same spin levels, i.e. the driven tunneling transitions. However,
in the presence of strong SOC, this active spatial motion 
triggers the spin flipping mechanism, which forms phase two where the
two same-spin initial state components decrease, whereas the other two with
opposite spin begin to grow. Finally, there is a third phase similar to the
first one, in which same-spin transitions take place. These final oscillations
help to ``settle'' the final state with the required spin orientation. 
Note that the electric field amplitudes are higher and oscillate faster in the
first and last phases of the process. Therefore, the spin-flip
transfer takes place during the relatively slow second phase, and occurs thanks to the
initial preparation of the faster field at the initial phase. Finally, some
faster motion is required in the last phase of the process in order to select
one specific final state in the subspace of states with the { desirable} spin direction.

{ { Although we have used given reference field amplitude $A_0$ for the starting-guess of the optimization algorithm; 
other values of $A_0$ could be considered.}
In a simple two-level Rabi picture, the population of the target state grows with the amplitude of a 
resonant monochromatic field. However, as already shown in Ref.~[\onlinecite{Khomitsky2012}], 
the population does not grow monotonically with increasing $A_0$. For the short time-intervals that 
we employ here, these target state populations are very small when using monochromatic fields, 
regardless of the $A_0$ amplitude used. Within the QOCT scheme, this $A_0$ is merely a starting 
value in the search algorithm; in fact, as discussed in the previous section, we have introduced 
a penalty in the target definition that leads to a preference of solutions with lower \emph{fluence}, 
i.e. average intensity over time. The reason for doing that is that we expect that the population of 
higher lying states, favored by higher amplitudes, would lead to faster 
decoherence.
Nevertheless, it is interesting to check how an optimization solution looks if we chose other values of $A_0$, since the solutions are not unique, 
and depend on the starting guesses. For that purpose we have performed an extra calculation for the driven spin-flip dynamics. In this case we have 
reproduced the conditions for Fig.~\ref{fig-2} with higher $A_{0}$ value, i.e. $A_0=3\times 10^2$ V$/$cm. The results (displayed 
in a Supplemental 
Material \footnote{See Supplemental Material for the results on stronger-field driving.}) show that the optimized driving field has a greater overall amplitude than that 
obtained in Fig.~\ref{fig-2}(a). The latter has resulted in the 
participation of higher energy orbital states (with the energies are around $U_0$ or higher), 
manifested in the high frequency 
oscillations in the dynamics of the observables.}

\subsection{Control of electron position}\label{ceap}
\begin{figure}
\centering
\includegraphics[trim=0cm 0cm 0cm 0cm,clip=true,width=0.425\textwidth]{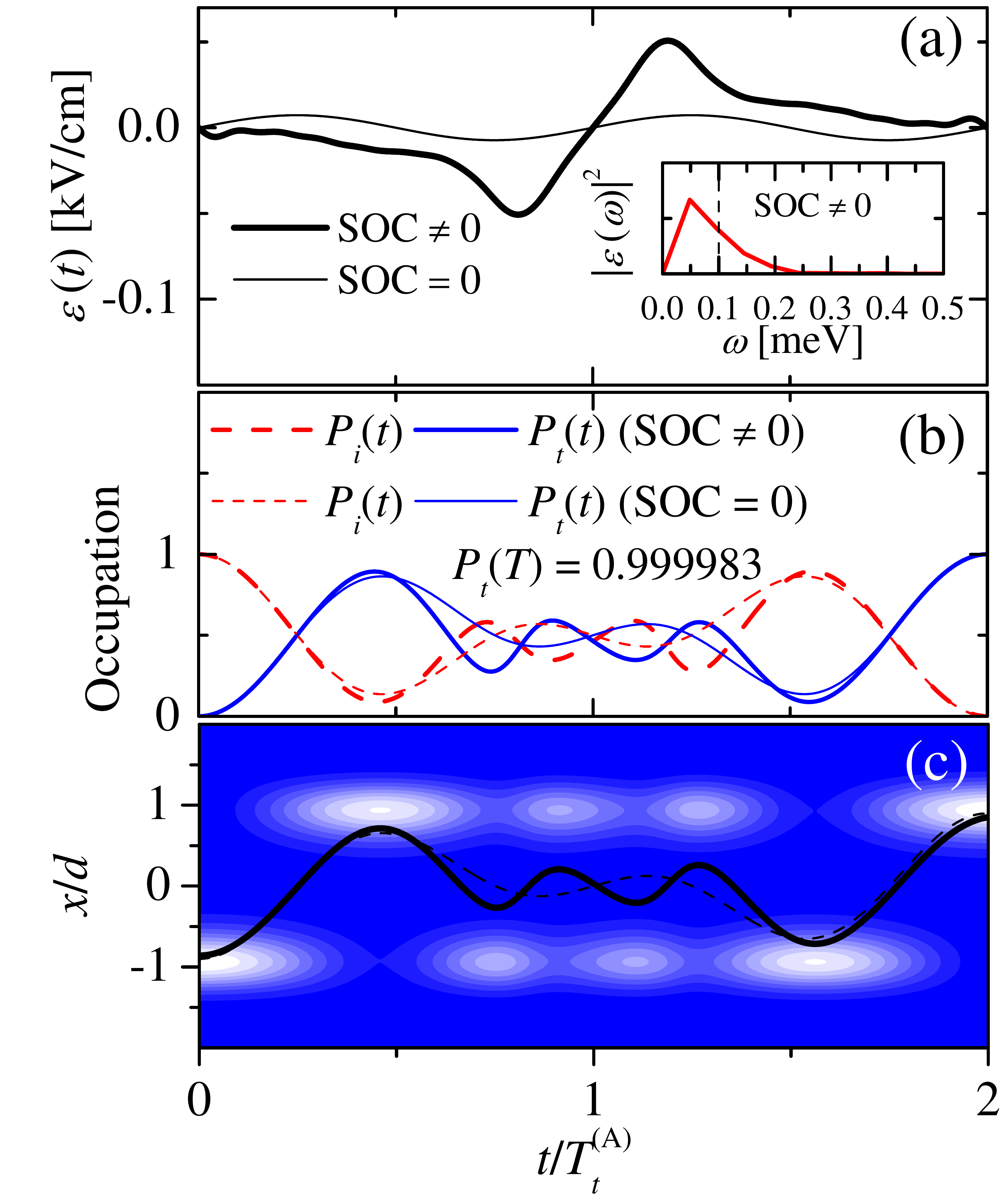}\\
\includegraphics[trim=0cm 0cm 0cm 0cm,clip=true,width=0.425\textwidth]{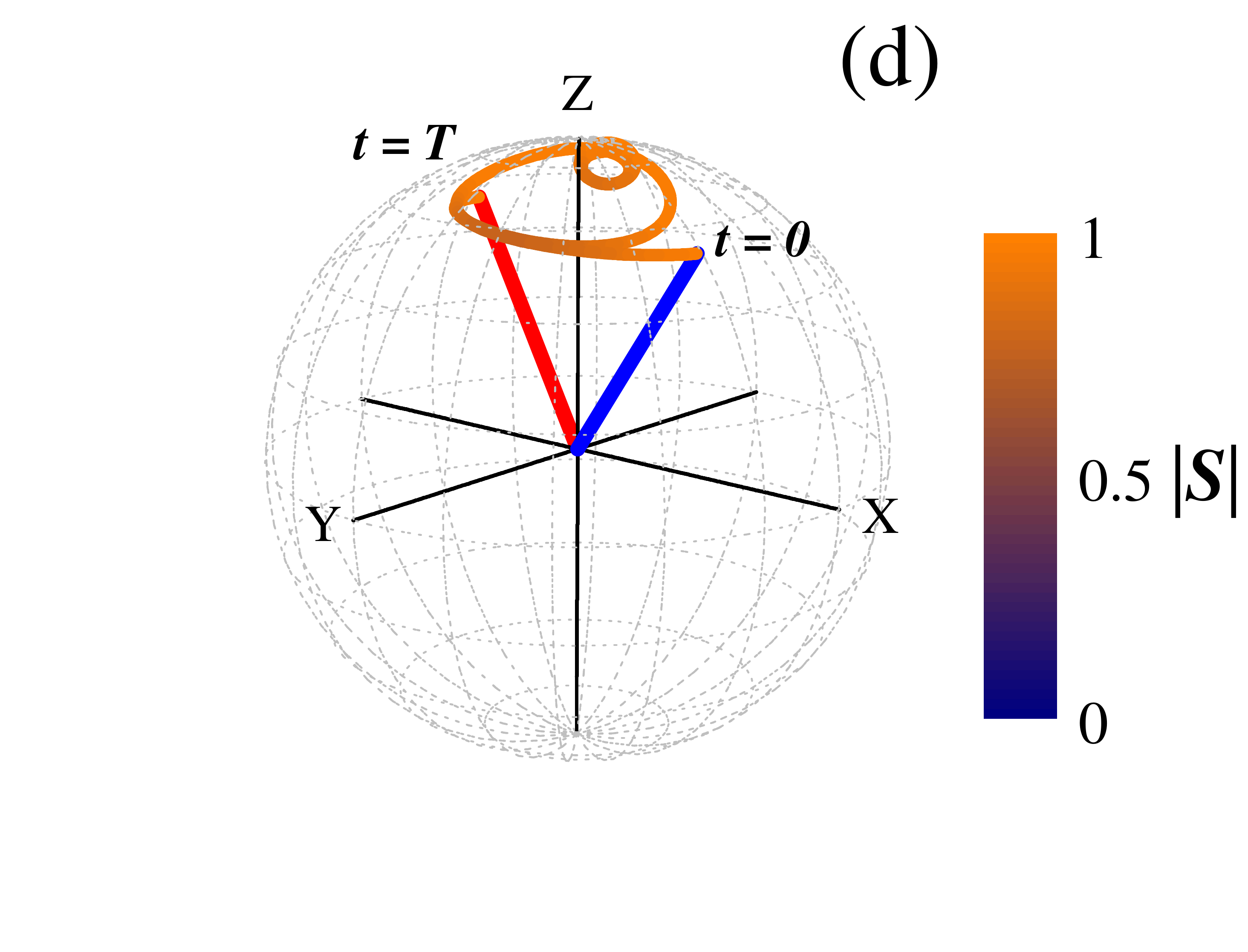}\\
\caption{(Color on-line) (a) Optimized driving field for a manipulation of electron position with $B=1.73$ T, both with and without SOC. 
In the inset, the power spectrum (arbitrary units) 
in frequency domain of the optimized pulse (with SOC) is shown; (b) time-dependence of the occupation of the initial and target states; 
(c) time-dependence of the charge distribution (with SOC). The thick black line is the time-dependent $\langle x\rangle$, while -- as comparison -- 
the thin dashed line correspond to the case without SOC; (d) trajectory of the tip of the spin vector, ${\bm S}(t)$ in the Bloch sphere (with SOC). 
The color scale represents the module of ${\bm S}$. The blue and red thick lines indicate the initial and final spin vectors, respectively.} \label{fig-4}
\end{figure}


In the example shown in this Section, the aim is manipulating the position of
the electron in the DQD system, keeping the spin $z$-component unchanged. For
this purpose, we begin with a state $\psi_i$ localized in one of the potential minima, 
and aim at moving it to the other one with conserved spin
orientation (the target state, $\psi_t$).  The initial and target states are not single eigenstates of the system, being each one of these a linear combination of two eigenstates:
\begin{eqnarray}\label{eq-17}
|\psi_i\rangle=\frac{1}{\sqrt{2}}\left( |\psi_0\rangle - i|\psi_2\rangle\right)\rightarrow
|\psi_t\rangle=\frac{1}{\sqrt{2}}\left( |\psi_0\rangle + i|\psi_2\rangle\right)\text{.}\nonumber\\
\end{eqnarray}
On the other hand, for the reference field we have chosen the driving frequency
$\omega_0$ corresponding to the splitting between the levels
$E_0$ and $E_2$: $\Delta E^{\rm (A)}_t=0.0967$ meV. The time for
one period of oscillation at this frequency is $T^{\rm (A)}_{t}=42.757$ ps.

Fig.~\ref{fig-4}(a) shows the pulse optimized to maximize the transition
$\psi_i\rightarrow\psi_t$. In order to assess the possible relevance of SOC
for this charge transfer transition, we have performed calculations both with
and without this term. (The Lyapunov-based control of the charge qubit in the absence
of SOC was recently analyzed in Ref.[\onlinecite{Cong2015}]). For these calculations, 
we have set the pulse length to $2T^{\rm (A)}_t$ since for shorter time intervals
it was impossible to reach target occupancy over $90$\%. Note that the power
spectrum of the field, presented in the inset, is centered around $\omega_0/2$. 
The optimized pulses allowed to nearly fully occupy the target state, as it can be seen in
Fig.~\ref{fig-4}(b). As for the optimized field, the dynamics of the occupancy of
states $\psi_i$ and $\psi_t$ shows odd symmetry with respect to the center of
the time-interval. The time-dependence of the populations of states $\psi_0$,
$\psi_1$, $\psi_2$ and $\psi_3$ (not shown here) reveals just a minor
contribution of the ``spin-down'' states ($\psi_1$ and $\psi_3$) around $t=T/2$. 
Fig.~\ref{fig-4}(c) shows the evolution of the charge distribution (including SOC) 
and $\langle x \rangle$ (for both, with and without SOC) as the electron is transferred 
between the minima. Finally, Fig.~\ref{fig-4}(d) shows the dynamics of the spin and a high-fidelity 
conservation of its $z$-component. One can see that the mean value of the 
electron spin performs a precession around the $z$-axis corresponding to
the dominating spin projection without reversing its sign during the 
whole time interval.

We conclude that the presence of SOC makes the charge manipulation slightly
harder, as the amplitude of the optimized field needs to be higher. 
In addition, the time-dependence of the field and the observables is more complex.

\subsection{Simultaneous spin-orbital control}\label{ssoc}

Next, in the final example, we seek to maximize the fidelity of a transition
that involves the transfer of the electron from one QD to the other, while reversing
the sign of $\langle\sigma_z\rangle$. Again, the system parameters and the
magnetic field that we have used are the same as in the previous example.

\begin{figure}
\centering
\includegraphics[trim=0cm 0cm 0cm 0cm,clip=true,width=0.425\textwidth]{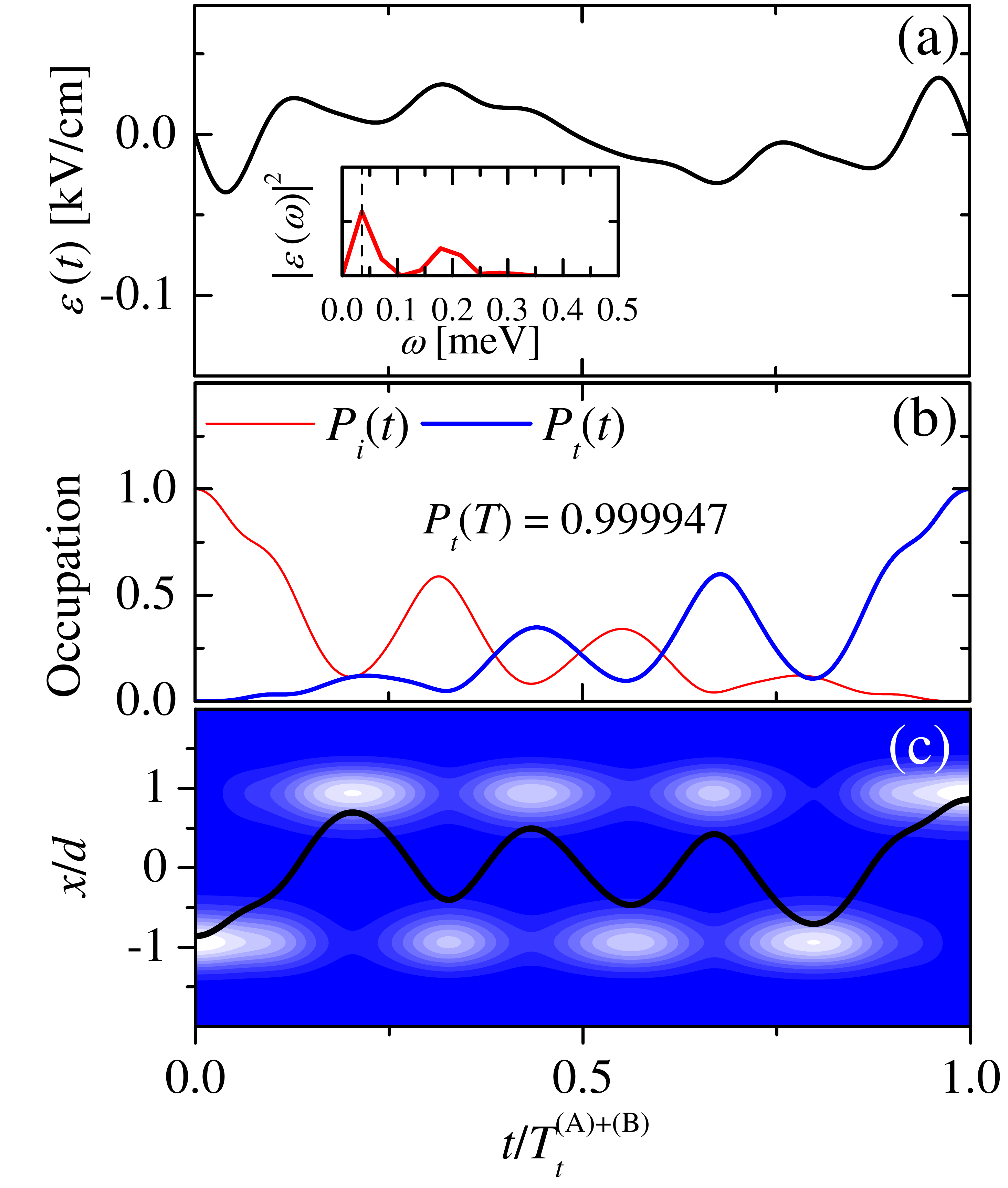}\\
\includegraphics[trim=0cm 0cm 0cm 0cm,clip=true,width=0.45\textwidth]{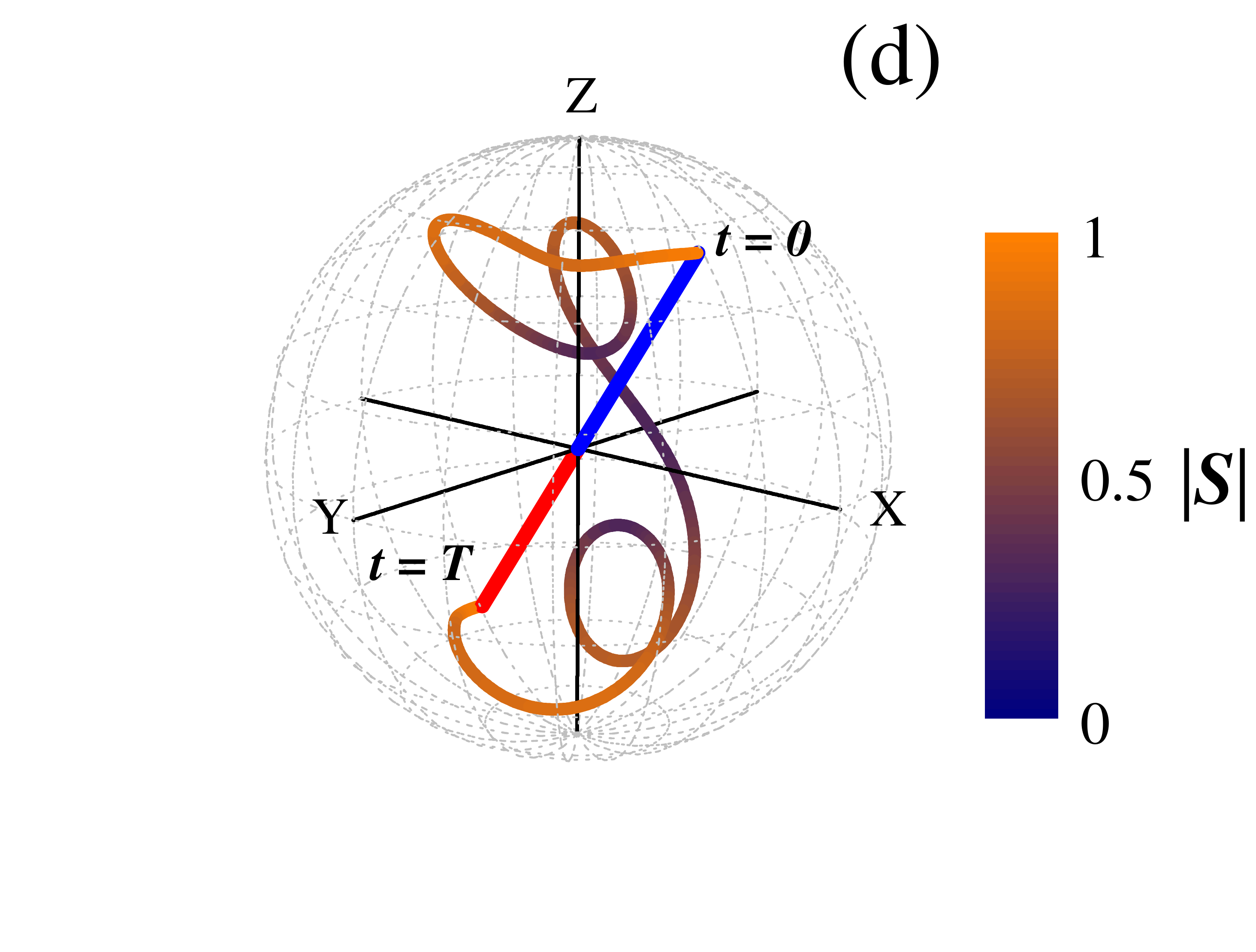}\\
\caption{(Color on-line) The same as Fig.~\ref{fig-4} but for a simultaneous 
spin-orbital manipulation with a magnetic field of $B=1.73$ T.} \label{fig-5}
\end{figure}

\begin{figure}
\centering
\includegraphics[trim=0cm 0cm 0cm 0cm,clip=true,width=0.425\textwidth]{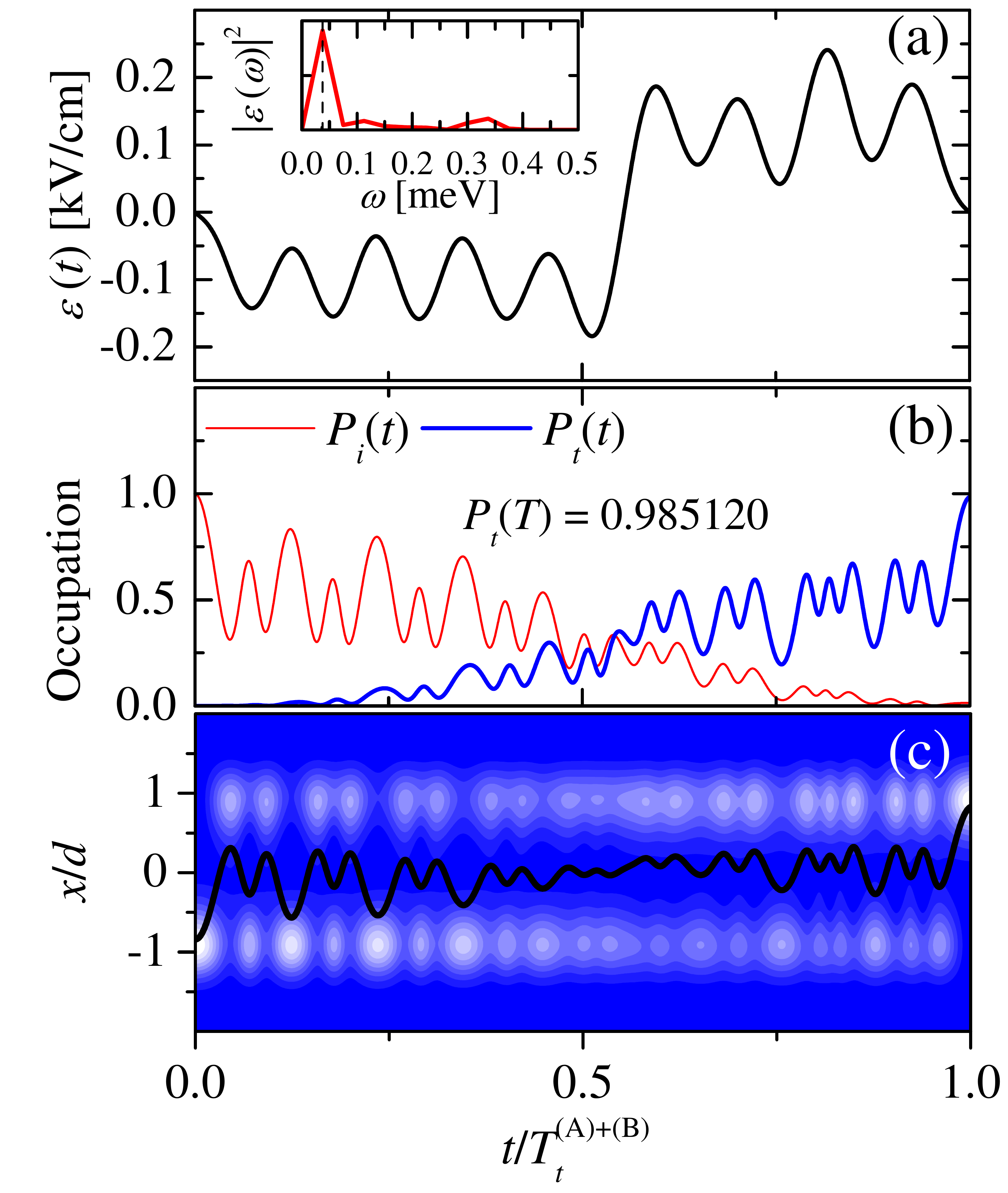}\\
\includegraphics[trim=0cm 0cm 0cm 0cm,clip=true,width=0.45\textwidth]{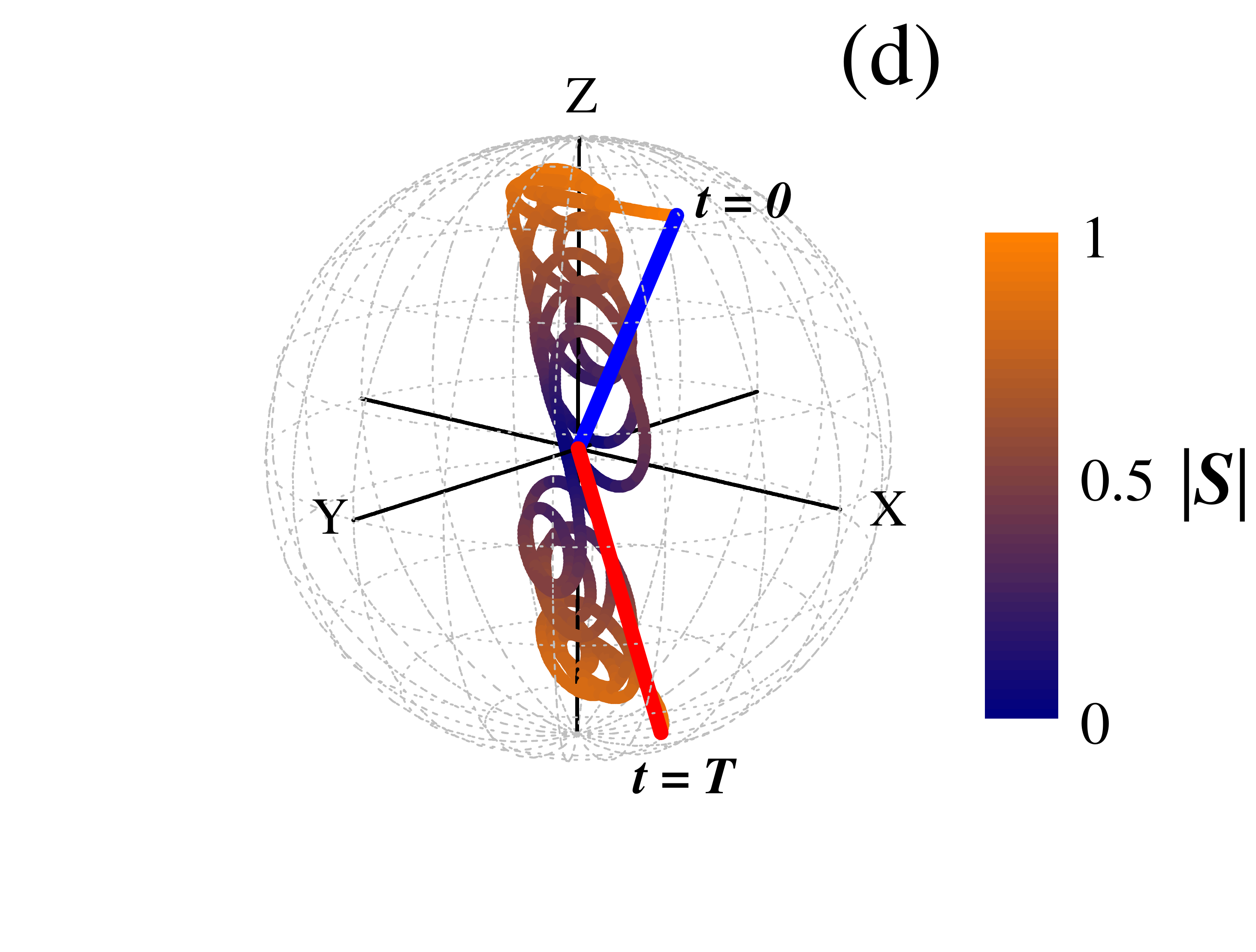}\\
\caption{(Color on-line) The same as Fig.~\ref{fig-5} but considering $U_0=5$ meV.} \label{fig-6}
\end{figure}


We start from the same initial state $\psi_i$ of the previous
example. The target, however, differs:
\begin{eqnarray}\label{eq-17}
|\psi_i\rangle=\frac{1}{\sqrt{2}}\left( |\psi_0\rangle - i|\psi_2\rangle\right)\rightarrow
|\psi_t\rangle=\frac{1}{\sqrt{2}}\left( |\psi_1\rangle - i|\psi_3\rangle\right)\text{.}\nonumber\\
\end{eqnarray}
In this state, the electron is located near the minimum at $x=d$ with negative $z$-component of spin.
The energy of this transition is $\Delta
E_t=\omega_0=(E_1+E_3)/2-(E_0+E_2)/2=(\Delta^{\rm (A)}_Z+\Delta^{\rm (B)}_Z)/2=0.0357$ meV 
and the associated period is $T^{\rm (A)+(B)}_t=116.06$ ps.

As in previous cases, the Figure~\ref{fig-5}(a) shows the pulse shaped to
maximize population of the target $\psi_t$. The length of the pulse in this
case is $T=T^{\rm (A)+(B)}_t$, which is enough to reach a target occupation of
$0.999947$, as it can be seen in Fig.~\ref{fig-5}(b). The population dynamics
of states $\psi_i$ and $\psi_t$ follows the relation
$|\langle\psi_t|\Psi(t)\rangle|^2=|\langle\psi_i|\Psi(-t)\rangle|^2$, noting
that in the central region both states are just slightly occupied. It is in
this region where the spin-flip transitions occur and almost 
the full occupation is due to the superposition of the $\psi_2$
(spin-up) and $\psi_1$ (spin-down) states (not shown here). Returning to
Fig.~\ref{fig-5}(a) we note in the inset that, as well as in
Fig.~\ref{fig-2}(a), the spectral component at the resonance frequency of the target transition,
$\omega_0=(\Delta^{\rm (A)}_Z+\Delta^{\rm (B)}_Z)/2$, remains the largest, dominating in the 
bimodal distribution. Finally, Fig.~\ref{fig-5}(c) shows the time-dependence of the charge distribution and the average electron position, whereas Fig.~\ref{fig-5}(d) presents the evolution of the spin vector. One can see a clear displacement of the electron from the left- to the right-QD 
with a simultaneous spin-flip. As was studied in Sec.~\ref{cdsfd}, the slow oscillatory behavior of the charge distribution between both QDs --in the central region of the time interval, reveals the region where spin-flip phase is occurring.

It is worth mentioning that in all the examples considered, the occupation of the higher energy states 
(above the lowest four) is practically negligible (below $10^{-3}$ in all cases). This not only happens for the dynamics generated by the optimized fields, but also for the dynamics produced by the monochromatic reference fields. The occupation of those higher energy states is difficult due to both the large energy gap between them and the lowest ones, and due to their small spatial overlap. In addition, it is important that the desired transitions can be obtained with relatively weak fields, also preventing occupation of the higher orbitals.

In all the previous examples we have fixed the structural parameters (distance between wells, barrier height, etc.). These parameters set a regime of interdot coupling that keeps the two well clearly separated, and separate the lower energy levels from the higher ones, although do not make them inaccessible. The variation of these parameters may lead to different dynamics, and we finish by discussing here these possibilities. First, by decreasing the interdot coupling, either by enlarging the distance or by increasing the barrier, we would get slower tunneling, but not qualitatively different results, as the higher energy levels would intervene even less. 

The opposite, that is an increase in the coupling, is more interesting. In order to study this realization, we analyzed another example with reduced potential barrier ($U_0=5$ meV) and the same distance between the QDs. Since the change in $U_0$ modifies the entire set of electron states, in order to obtain the same final configuration as in the previous case the target state must be set as $|\psi_t\rangle=\left(|\psi_1\rangle+i|\psi_3\rangle\right)/\sqrt{2}$. Obviously, the change in $U_0$ also modifies the tunneling 
splitting, $\Delta E_t$ and the values of $\omega_0$ ($0.0375$ meV), $T_t^{\text{(A)+(B)}}$ ($110.32$ ps) and $A_0$ ($7.5\times 10^1$ V$/$cm).

Fig.~\ref{fig-6}(a) shows the optimized field obtained in this simulation, where the presence of higher frequency components of larger amplitude is well-seen. Nevertheless, the resonant frequency of the target transition is still the main component, as can be seen in the inset. 
The fact that the optimized field includes these higher frequencies is a sign of the increased presence of transitions involving states above the lowest four. This is logical given that, as $U_0$ decreases, the gap between these lowest states and the higher ones decreases too, so that the transient 
occupations of the latter becomes more probable. In the previous case [Fig.~\ref{fig-5}(a)], the energy gap prevents the optimization algorithm from making use of those higher energy states, avoiding the need of electric field oscillations of greater amplitude. On the contrary, in this case [Fig-~\ref{fig-6}(a)] 
this gap is smaller, the higher energy states more accessible, and therefore the optimization algorithm uses them.

The presence of higher energy transitions during the dynamics is also reflected in the 
evolution of the observables (see Figs.~\ref{fig-6}(b),~\ref{fig-6}(c) and ~\ref{fig-6}(d)). 
Note the higher frequency oscillations in the three cases compared with those observed in Fig.~\ref{fig-5}. In particular, in Fig.~\ref{fig-6}(c), note that the charge density oscillates much faster and with smaller amplitudes (the charge is almost equally distributed in both QDs), in comparison with Fig.~\ref{fig-5}(c). Note also that the charge distribution penetrates more deeply into the barrier region throughout all the dynamics.

In Sec.~\ref{cdsfd}, we referred to the fact that the QOCT scheme aims at to minimizing  driving field fluence. For the present case we have reduced the amplitude of the reference driving field. However, it was interesting to see that even if we keep the original amplitude ($A_0=1.5\times 10^2$ V$/$cm), the results (not shown here) are quantitatively similar to those shown in Fig.~\ref{fig-6}.

\section{Conclusions}\label{conclusions}

This paper has shown how to optimally control, using the
spin-orbit coupling, the electron localization
and the simultaneous spin dynamics in single-electron nanowire-based
double quantum dots by electric means. 
The manipulation is fast (of the order of 0.1 ns, much shorter than the  decoherence time induced by the unavoidable hyperfine interactions \cite{Merkulov2002}) 
if the electric pulses are properly shaped, which we have achieved with the help of quantum optimal control theory. 
The dynamics of these systems had been known to be complex (and to some extent
surprising) as it was found~\cite{Khomitsky2012} that the Rabi spin oscillations
frequency does not grow monotonously with the electric field amplitude, but
rather exhibits an unexpected nonlinear behavior. Due to this fact, the
use of monochromatic radiation results in rather slow spin-flip transitions. In addition and for the same reason, it is not useful to simply increase the field amplitude. This is a challenge if these systems are to be used in spintronics devices. Here we have shown how the problem can be solved by making use of control techniques to design more complex electric pulses. The obtained shape of the pulse is relatively simple and includes less than 10 main spectral components, just simplifying the pulse generation.   

In this paper we concentrated on the systems with the parameters corresponding to the spectrum of the optimal pulses in the sub-THz domain. It is possible to modify the double quantum dot parameters such that the splitting of the levels, and, correspondingly, the spectral range of the pulses will 
become of the order 1 GHz. This is a better frequency scale for conventional semiconductor-based electronics and involves the use of lower magnetic fields. 
In this case the pulse design is based on the same approach as we applied here, and the results will be similar to the results obtained here.

\begin{acknowledgments}
J.A.B. and A.C. are grateful for the support offered by the European
Community FP7 through the CRONOS project, grant agreement No. 280879, and
the Spanish Grant FIS2013-46159-C2-2-P. D.V.K. is supported by the RFBR Grant No. 15-02-04028-a, 
and by the University of Nizhny Novgorod and the Russian Academy of Science Joint Laboratories Project.
This work of E.Y.S. was supported by the University of Basque
Country UPV/EHU under program UFI 11/55, Spanish MEC (FIS2012-36673-C03-01), 
and Grupos Consolidados UPV/EHU del Gobierno Vasco (IT-472-10).
\end{acknowledgments}

\bibliographystyle{apsrev}
\bibliography{bib_dqd_PRB}

\end{document}